\documentclass[amstex,epsf,amssymb,usenatbib]{mn2e}
\usepackage{epsf,graphics}
\usepackage{etex}
\reserveinserts{30}
\usepackage{amsmath,amssymb, natbib}
\usepackage{rotating,times,pictex,graphicx,latexsym,color,longtable}
\title{About the nature of Mercer\,14} 

\author[Froebrich \& Ioannidis]{D.~Froebrich$^{1}$\thanks{E-mail:
df@star.kent.ac.uk}, G.~Ioannidis$^{1}$\thanks{E-mail: gi8@kent.ac.uk}\\ $^1$
Centre for Astrophysics and Planetary Science, University of Kent, Canterbury,
CT2 7NH, UK } 

\begin{document}

\date{Received sooner; accepted later}
\pagerange{\pageref{firstpage}--\pageref{lastpage}} \pubyear{2007}
\maketitle

\label{firstpage}

\begin{abstract}

We used UKIRT near infrared (NIR) broad band {\it JHK} photometry, narrow band
imaging of the 1-0\,S(1) molecular hydrogen emission line and mid infrared {\it
Spitzer} IRAC data to investigate the nature of the young cluster Mercer\,14.
Foreground star counts in decontaminated NIR photometry and a comparison with
the Besancon Galaxy Model are performed to estimate the cluster distance. This
method yields a distance of 2.5\,kpc with an uncertainty of about 10\,\% and can
be applied to other young and embedded clusters. Mercer\,14 shows clear signs of
ongoing star formation with several detected molecular hydrogen outflows, a high
fraction of infrared excess sources and an association to a small gas and dust
cloud. Hence, the cluster is less than 4\,Myrs old and has a line of sight
extinction of A$_K$\,=\,0.8\,mag. Based on the most massive cluster members we
find that Mercer\,14 is an intermediate mass cluster with about 500\,M$_\odot$.

\end{abstract}

\begin{keywords}
Galaxy: open clusters, individual; ISM: jets and outflows; stars: formation
\end{keywords}

\section{Introduction}

Understanding the formation of stars is one of the key topics of current
astrophysical research. It occurs in embedded clusters within giant molecular
clouds. A large fraction of all young stars are formed in such clusters (e.g.
Lada \& Lada \citeyearpar{2003ARA&A..41...57L}). Low star formation efficiency, as well
as feedback (radiation, winds, outflows) from young and massive stars leads to
the disruption of many of those clusters in the first 10\,Myrs of their life and
generates the field star population. Thus, in order to understand the formation
of stellar clusters one has to investigate the youngest (up to a few Myrs old)
objects. Preferably a large number of objects should be investigated to account
for environmental influences, but detailed investigations of individual objects
are vital to accurately determine their physical parameters such as age,
distance and line of sight reddening. 

Investigations of these young embedded clusters are hampered by the high column
density material in their vicinity. Recent years have seen advances of large
scale near and mid-infrared surveys (2 Micron All Sky Survey (2MASS, Skrutskie
et al. \citeyearpar{2006AJ....131.1163S}); UK Infrared Deep Sky Survey (UKIDSS,
Lawrence et al. \citeyearpar{2007MNRAS.379.1599L}) Galactic Plane Survey (GPS, Lucas et
al. \citeyearpar{2008MNRAS.391..136L}); GLIMPSE, Benjamin et al.
\citeyearpar{2003PASP..115..953B}) which allow us to discover and characterise the
properties of a variety of these young embedded clusters at wavelengths less
influenced by interstellar extinction. A large number of so far unknown stellar
clusters in the Galaxy have been found in those surveys (e.g. Bica et al.
\citeyearpar{2003A&A...404..223B}. Mercer et al. \citeyearpar{2005ApJ...635..560M}, Froebrich
et al. \citeyearpar{2007MNRAS.374..399F}, to name just a few). 

In particular the list by Mercer et al. \citeyearpar{2005ApJ...635..560M}, based on
mid-infrared {\it Spitzer} data, contains a number of embedded young clusters
hidden from view in shorter wavelengths surveys. One of these, Mercer\,14, has
so far not attracted any attention besides the discovery paper. We aim to
determined the nature of this cluster mainly by means of near infrared {\it JHK}
photometry obtained from the UKIDSS GPS and the UK Infrared Telescope (UKIRT)
Widefield Infrared Survey for H$_2$ (UWISH2, Froebrich et al.
\citeyearpar{2011MNRAS.413..480F}) which is using a narrow band filter centred on the
1-0\,S(1) emission line of molecular hydrogen at 2.122\,$\mu$m. 

We briefly discuss the data used in our paper in Sect.\,\ref{data}. Our
analysis, in particular the results of the decontamination of the photometry,
the cluster distance determination and the identification of molecular outflows
from young stars in the cluster are discussed in Sect.\,\ref{method}. We then
put forward our conclusions in Sect.\,\ref{conclusions}.

 %
 %

\section{Data and Analysis}\label{data}

\subsection{Near infrared UKIRT WFCAM data}

We obtained near infrared imaging data in the 1-0\,S(1) line of molecular
hydrogen using the Wide Field Camera (WFCAM, Casali et al.
\citeyearpar{2007A&A...467..777C}) at UKIRT. The data are part of the UWISH2 survey
(Froebrich et al. \citeyearpar{2011MNRAS.413..480F}) and the H$_2$ images are taken
with a per pixel integration time of 720\,s and under very good seeing
conditions. The full width half maximum of the stellar point spread function is
0.7\arcsec and the 5\,$\sigma$ point source detection limit is about 18\,mag.
Our narrow band data were taken on the 30th of July in 2009.

In order to perform the continuum subtraction of our narrow band images we
further utilised the UKIDSS data in the near infrared {\it JHK} bands taken with
the same telescope and instrumental setup as part of the GPS. This broad band
data was taken on the 1st of June in 2006 in all three filters. All GPS
data used was taken from data release 7.

All NIR (UWISH2 and GPS) data reduction and photometry are performed by the
Cambridge Astronomical Survey Unit. Reduced images and photometry tables are
available via the Wide Field Astronomy Unit and are downloaded from the WFCAM
Science Archive (Hambly et al. \citeyearpar{2008MNRAS.384..637H}). The basic data
reduction procedures applied are described in Dye et al.
\citeyearpar{2006MNRAS.372.1227D}. Calibration (photometric as well astrometric) is
performed using 2MASS (Skrutskie et al. \citeyearpar{2006AJ....131.1163S}) and the
details are described in Hodgkin et al. \citeyearpar{2009MNRAS.394..675H}. All NIR
magnitudes used in this paper are quoted for {\tt AperMag3}, i.e. a 2\arcsec\
aperture, the standard for GPS point sources (see Lucas et al.
\citeyearpar{2008MNRAS.391..136L} for details).

\subsection{Mid infrared {\it Spitzer} IRAC data}
\label{spphot}

We also downloaded IRAC images for all filters and point source photometric
catalogues of the region that have been obtained by {\it Spitzer} as part of the
GLIMPSE Survey (Benjamin et al. \citeyearpar{2003PASP..115..953B}). All data is part of
the final release of enhanced data products for GLIMPSE. Except two, all of the
potential outflow driving sources for which we require photometry are extended.
Hence, they are not part of the GLIMPSE point source catalogue. We thus
performed aperture photometry on the IRAC images.

We defined apertures encompassing all the flux from the extended sources, as
well as apertures around some bright point sources (e.g. obj.\,C, see
Table\,\ref{sources}). Since the cluster is embedded in partly dense material,
the background flux for each source has been determined in an aperture as close
to the source as possible. Fluxes above the background were then converted into
magnitudes and calibrated using the bright source obj.\,C, whose magnitudes are
taken from the GLIMPSE point source catalogue. Uncertainties introduced due to
background estimation and the choise of aperture are about 0.1\,mag for the two
short wavelengths filter. For the longer wavelengths we estimate a larger error
of about 0.2\,mag, since the background emission is stronger (in particular in
the 8\,$\mu$m filter) and spatially variable on small scales. In conjunction
with the GLIMPSE calibration uncertainty of 0.1\,mag to 0.2\,mag our photometric
uncertainties are 0.2\,--\,0.3\,mag with the larger value valid for the longer
wavelengths filters.

\section{Analysis and discussion of Mercer\,14}\label{method}

\subsection{General}
 
The cluster Mercer\,14 has been discovered by Mercer et al.
\citeyearpar{2005ApJ...635..560M} during their search for embedded star clusters
utilising GLIMPSE data. The cluster is embedded in a small, mid-infrared bright
molecular cloud. It has, except in the discovery paper, not attracted any
attention nor have its parameters be determined. The near infrared {\it JHK}
images give the impression of a spherically symmetric accumulation of stars,
clearly visible at near infrared wavelengths. The central coordinates,
determined visually are at 18:58:05.8 $+$01:36:34, in perfect agreement with the
coordinates given in Mercer at al. \citeyearpar{2005ApJ...635..560M}. Visually the
radius of the cluster is about half an arcminute. 

\begin{figure*}
\centering
\includegraphics[width=10cm]{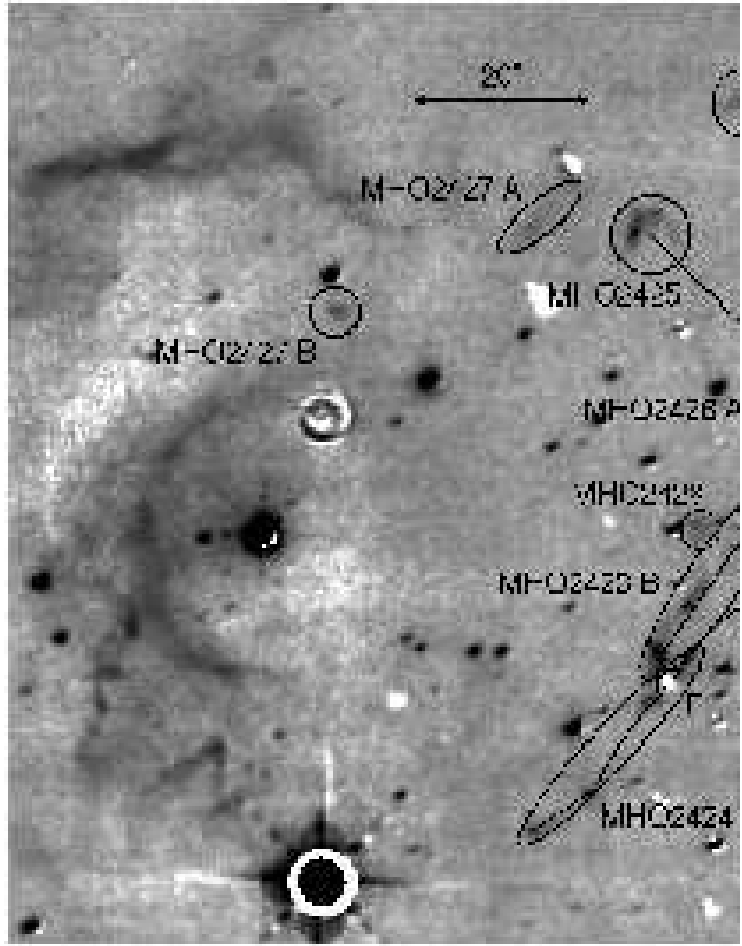}

\caption{\label{h2k_image} Gray scale representation of the H$_2$-K difference
image of the region containing outflows near Mercer\,14. The image is centred at
RA\,=\,18:58:06.6 and DEC\,=\,+01:37:03 (J2000). North is up, East to the left.
Circles and ellipses mark the MHOs discussed in the text. Their numbers are
added as labels. Squares indicate the five high probability driving sources and
the solid lines mark some of the discussed flows. North is to the top and East
to the left.} 

\end{figure*}

Our continuum subtracted molecular hydrogen images of the cluster vicinity show
that there are a number of H$_2$ flows and features, mostly just to the
North-North-East of the cluster (see Fig.\,\ref{h2k_image}). If they are
associated with the cluster itself, then this would indicate that Mercer\,14 is
a young stellar cluster, with partly still ongoing star formation. At the
northern edge of the cluster there is a very red, fuzzy nebulosity visible in
the {\it JHK} colour images. It coincides with one of the IRAC detected Extended
Green Objects (EGOs; EGO\,G035.13-0.74; Cyganowski et al.
\citeyearpar{2008AJ....136.2391C}; Obj.\,A in Table\,\ref{sources}). Only 4\arcsec\ to
the north-east of this, we find a smaller red reflection nebula, a fainer EGO
not included in the list by Cyganowski et al. \citeyearpar{2008AJ....136.2391C} 
(Obj.\,B). About 75" to the North-East-East of the cluster centre we detect a
bright star (K\,=\,9.5\,mag in 2MASS) in the centre of a small HII region. This
object is actually listed as a Planetary Nebula (PN\,G035.1-00.7) in the SIMBAD
database. The NIR images, however, clearly suggest that this object is an HII
region (see a more detailed discussion in Sect.\,\ref{hiiregion}).

\begin{figure*}
\centering
\includegraphics[width=8.5cm]{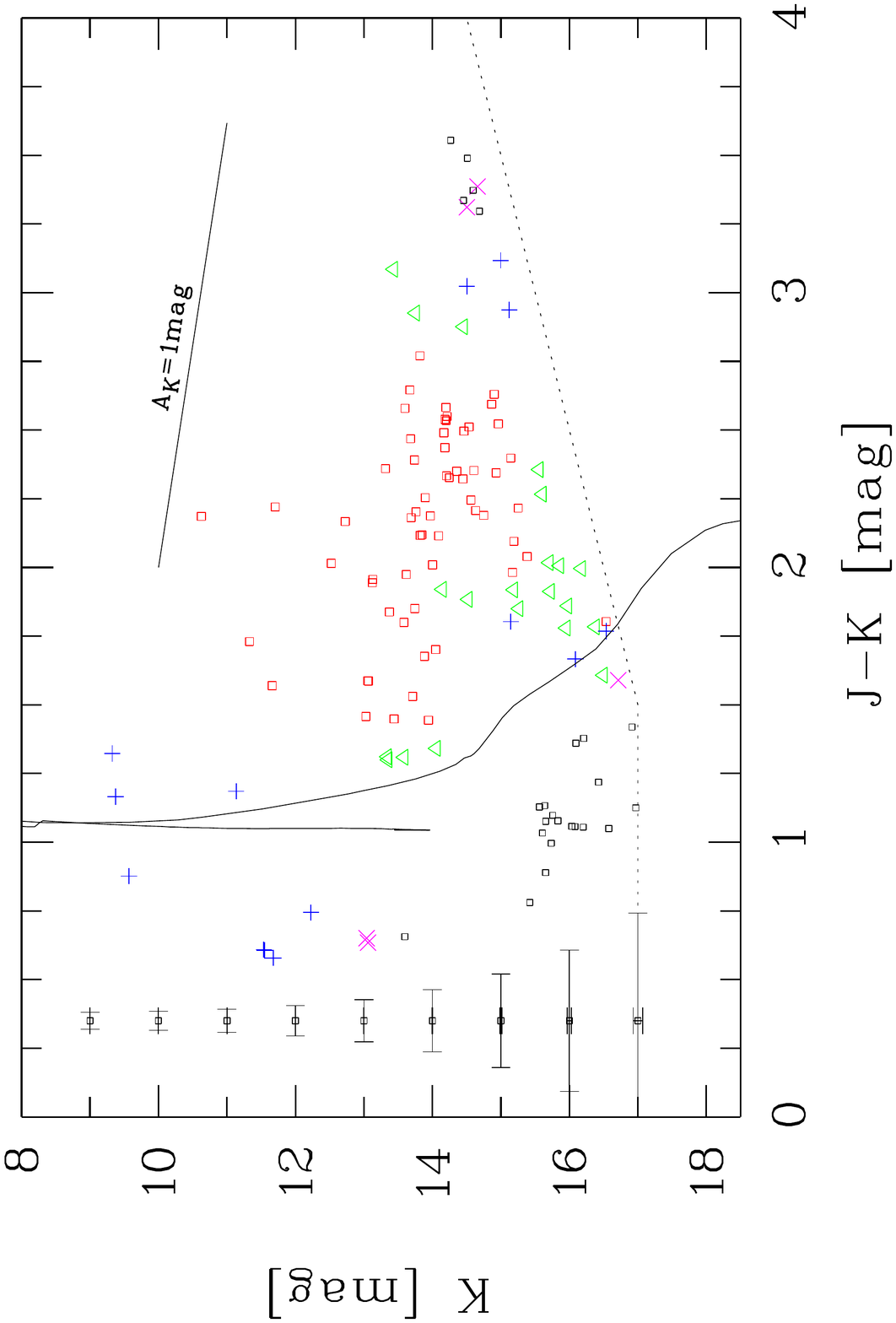}\hfill
\includegraphics[width=8.5cm]{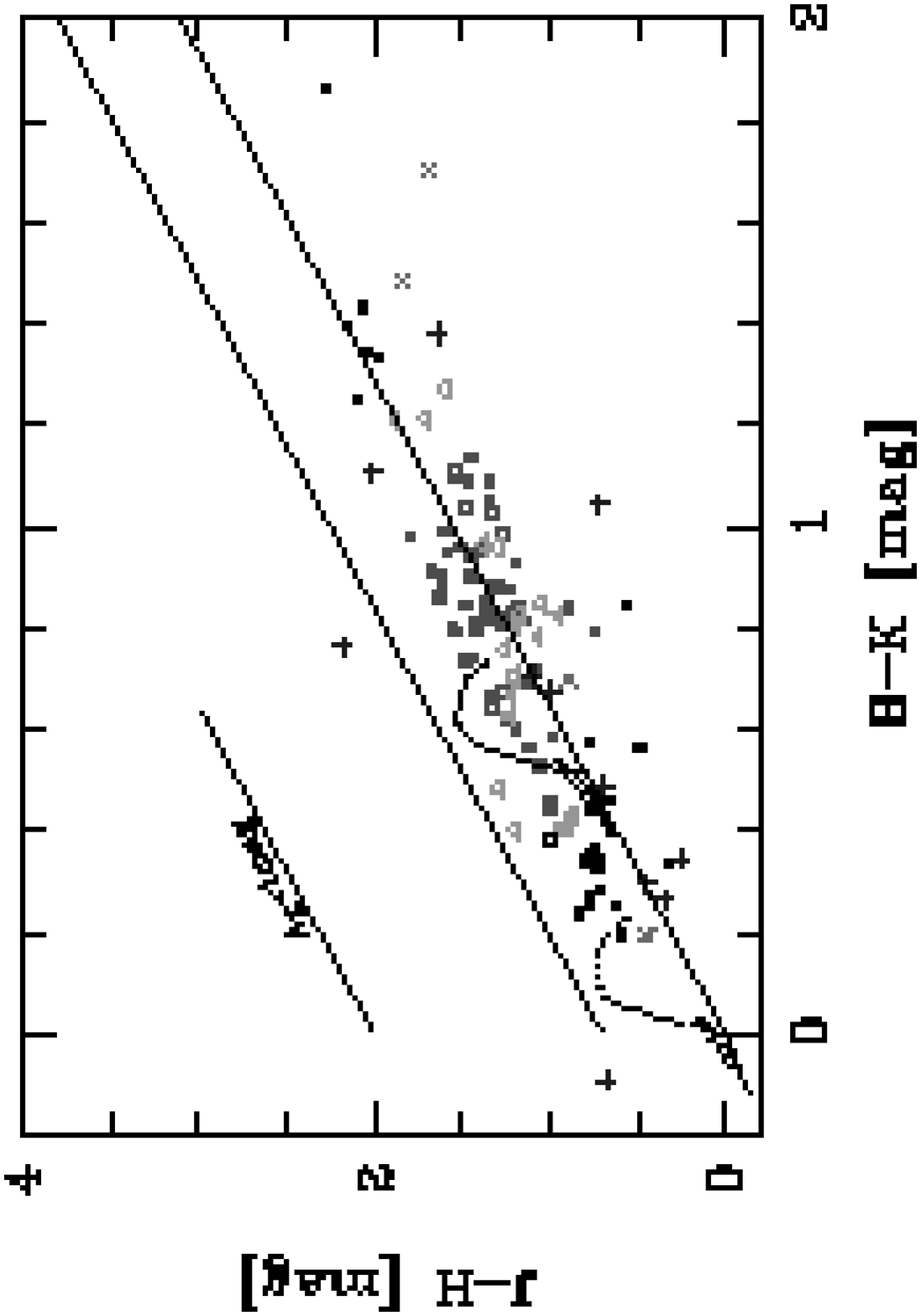}

\caption{\label{mercer14_decon} Decontaminated colour-magnitude (left panel) and
colour-colour (right panel) diagram of the cluster Mercer\,14. Only objects in
the GPS with {\it pstar} greater than 0.9 are included. Cluster membership
probabilities $P_{\rm ccm}$ are indicated by different symbols: above 80\,\% red
square; 60--80\,\% green triangle; 40--60\,\% blue plus sign; 20--40\% magenta
cross; below 20\,\% black dot. The $P_{\rm ccm}$ probabilities are based on the
15th nearest neighbour in the colour-colour-magnitude space. The over plotted
isochrone (solid line) has an age of 4\,Myrs, a distance of 2.5\,kpc and a
K-band extinction of 0.8\,mag. The dashed isochrone in the right panel has the
same properties but is un-reddened. The dotted line in the left panel indicates
the detection limits in the cluster area. The solid diagonal lines show the
reddening band for photospheres based on our adopted extinction law.}

\end{figure*}

\subsection{NIR cluster decontamination}
\label{decon}

In order to analyse the cluster properly, we need to decontaminate the NIR
photometry from fore and background objects. The decontamination is based on the
method established by Bonatto \& Bica \citeyearpar{2007MNRAS.377.1301B}. But we use the
improvements introduced by Froebrich et al. \citeyearpar{2010MNRAS.409.1281F}. We
download the near infrared photometry from the GPS within 3\arcmin\ around the
cluster centre. Only stars with a detection above the local completeness limit
(J\,=\,18.5\,mag, H\,=\,17.5\,mag, K\,=\,17.0\,mag) in each filter, photometric
uncertainties of less than 0.1\,mag, and with a {\it pstar} value of more than
0.9 are considered (i.e. only star like objects, see Lucas et al.
\citeyearpar{2008MNRAS.391..136L}).

We define as 'cluster area' everything within 0.9\arcmin\ from the cluster
centre. As 'control region' we use all stars between 2\arcmin\ and 3\arcmin\
from the central coordinates. The calculation of the $P_{\rm ccm}$ membership
probabilities for the individual stars in the cluster area is based on the
15$^{th}$ nearest neighbour in the colour-colour-magnitude space (see Froebrich
et al. \citeyearpar{2010MNRAS.409.1281F} for details).

The result of the procedure is shown in the decontaminated colour-magnitude
(CMD) and colour-colour (CCD) diagrams in Fig.\,\ref{mercer14_decon} together
with a cluster isochrone as a guide to the cluster sequence (see
Sect.\,\ref{clusterproperties}). There is a large fraction of stars with $P_{\rm
ccm}$ membership probabilities of more than 60\,\%, as well as possible fore or
background objects not related to the cluster.

\subsection{Distance determination}

There are a variety of ways to determine the distance to Mercer\,14 using the
available data set. These are: i) Using an object associated to the cluster or
its parental molecular cloud which has a known distance; ii) Perform isochrone
fitting to the cluster's decontaminated CMD and CCD; iii) Estimate the number of
foreground stars to the cluster or its parental molecular cloud and compare this
to the Besancon Galaxy model by Robin et al. \citeyearpar{2003A&A...409..523R}. 

i) A search in the vicinity of Mercer\,14 for objects of known distance did not
reveal any usable object. The object nearest to the cluster with a reliable
distance estimate is the massive star forming object G35.2N. Uruquart et al.
\citeyearpar{2011MNRAS.410.1237U} place it at a distance of 2.3\,kpc. Considering the
projected separation of about 4.3\arcmin\ (or about 3\,pc at this distance), it
is not entirely clear if the cluster and G35.2N are physically related. The
8\,$\mu$m GLIMPSE image of the region also reveals that Mercer\,14 is just
surrounded by a small cloud with no obvious connection to G35.2N.

ii) Given the low age and large number of pre-main sequence stars in the cluster
(indicated by the presence of molecular outflows, as well as the isochrone fit,
see below) it is extremely difficult to simultaneously fit the distance and age
of the cluster to satisfactory accuracy. Rather the distance needs to be known
in order to determine the approximate age and reddening to the cluster from
photometry alone. Thus, an estimate of the number of stars foreground to the
cluster or the parental molecular cloud can be determined and compared to the
Besancon Galaxy model (Robin et al. \citeyearpar{2003A&A...409..523R}) in order to
estimate the distance. It has been shown e.g. by Scholz et al.
\citeyearpar{2010MNRAS.406..505S} that such an approach can give a reasonably accurate
distance estimate. 

iii) As a first possibility we count the number of blue 
(J\,-\,K\,$<$\,2.0\,mag) stars per unit area across the parental molecular cloud
of the cluster. We used six fields near the cluster to count foreground stars
based on their colours. They are spread at different distances from the cluster
(2\arcmin\,--\,4\arcmin). Fields too close to the cluster might contain cluster
members, contaminating the analysis. On the other hand, fields too far away
might show a too low extinction to successfully distinguish stars foreground or
background to the cloud. One might even measure the distance to a cloud not
related to the cluster at all, given its position in the Galactic Plane and the
possibility of overlapping clouds along the line of sight. The area where
foreground stars are counted should be as large as possible to minimise
statistical errors. We choose one square arcminute sized fields. The number of
foreground stars in them varies between 13 and 22, with an average of 18 stars
per square arcminute. 

We then estimated the photometric properties of the near infrared data set
obtained from the GPS (completeness limit and photometric errors as a function
of magnitude). A model photometric catalogue for 10 square arcminutes (we chose
a larger area to improve the accuracy of the model) towards the cluster position
and with the above determined restrictions for the {\it JHK} photometry was
downloaded from the Besancon galaxy model
webpage\footnote{http://model.obs-besancon.fr/}. A comparison of the measured
number of foreground stars with the model predictions leads to a distance
estimate. The model stars are sorted by distance. We find the star in the list
for which all stars closer than it would give a model star density equal to our
measured foreground star density. The distance to this model star is then used
as the distance to the cloud. By considering the variations of the forground
star density we obtain a range of distances for the cloud of 2.1\,kpc to
2.8\,kpc with an average of 2.5\,kpc.

The second possibility is to determine the number of stars per unit area that
are foreground to the cluster itself. This is based on the decontaminated CMD of
the cluster (see Sect.\,\ref{decon}). The colour of most of the high probability
cluster members is $J-K =$\,2\,mag or redder. Hence all stars bluer than this
are potentially foreground stars. The total number of stars bluer than $J-K
=$\,2\,mag is 71 within a radius of 0.9\arcmin\ around the cluster centre. If we
weight the contribution of each star by the probability that it is not a cluster
member ($1 - P_{\rm ccm}$) then there are 17 stars per square arcminute
foreground to the cluster. This corresponds to a distance of 2.5\,kpc, in
perfect agreement to the star counting estimates using the fields in the
parental molecular cloud. This distance has been determined using the nearest 15
neighbours in colour-colour-magnitude space for the decontamination. When using
the nearest 10 neighbours the distance estimate is about 2.0\,kpc, for the 20th
nearest neighbours we obtain 2.8\,kpc. Thus, including the uncertainties, the
value also agrees with the distance of G35.2N. Hence, Mercer\,14 could indeed be
part of the same star forming complex as the massive YSOs in G35.2N.

We use 2.5\,kpc as distance to Mercer\,14 throughout the paper.

\subsection{Outflows and driving sources}

As can be seen in Fig.\,\ref{h2k_image}, there are a number of molecular
hydrogen outflow features just to the North-East of the cluster (labeled
MHO\,2423\,--\,MHO\,2428). In the following we will describe their properties
and discuss possible driving sources. This discussion will be focused on the
most obvious driving source candidates listed in Table\,\ref{sources} and
labeled by letters A\,--\,E in Fig.\,\ref{h2k_image}. There are a larger number
of other objects in the cluster region that show near infrared excess. About one
in three of the high probability cluster members show K-band excess emission
(objects to the right of the reddening band, see right panel of
Fig.\,\ref{mercer14_decon}). Each of these could in principle drive an outflow
and some of them might be the source for H$_2$ emission features. Due to their
large numbers and without further information such as proper motions of the
emission knots we cannot include them into our considerations. 

The numbering and naming of the molecular hydrogen emission line objects (MHOs)
has been done in accordance with the convention set out by Davis et al.
\citeyearpar{2010A&A...511A..24D}. In the Appendix (online version only) we present two
higher resolution continuum subtracted H$_2$-K images (Figs.\,A1 and A2) with
detailed labels for each identified MHO. None of the outflow features is visible
in the H$\alpha$ images taken by the IPHAS survey (Drew et al.
\citeyearpar{2005MNRAS.362..753D}). This is most likely caused by the high optical
extinction towards the cluster (see below).


A summary of the MHO positions, fluxes and possible driving sources is given in
Table\,\ref{mhotab}.

\subsubsection*{MHO\,2423}


Objects MHO\,2423\,A and MHO\,2423\,B seem to form two sides of a symmetric
bipolar outflow with a position angle of about 150$^\circ$. It has a length of
1\arcmin, which corresponds to about 0.75\,pc at the adopted distance. Given the
symmetry of the flow, the driving source should be situated on the axis defined
by the two features and be located roughly half way between the two ends. There
is no obvious bright source detectable in the NIR data. One of the two red and
extended sources (obj.\,B in Table\,\ref{sources} -- the faint EGO) is situated
just off-axis. There is, however, a relatively bright and red {\it Spitzer}
source (obj.\,D, undetected in the NIR images) closer to the axis. Based on its
position and red ([3.6-4.5]\,=\,3.1\,mag) colour it is the most likely source
for the flow. 

\subsubsection*{MHO\,2424}


The identification of the source(s) of MHO\,2424\,A and MHO\,2424\,B is not so
straight forward. Both could be driven by the bright EGO (obj.\,A). However, the
small chain of three knots seems not to be aligned with this extended source. It
also could be driven by the bright IRAC source (obj.\,C) which is completely
invisible in the NIR images. In this case the flow would be one-sided and
23\arcsec\ (0.3\,pc) long. Given that we do not detect any counter flow and that
there is plenty of material in the vicinity of the cluster, this does not seem a
likely explanation.

Another possibility is that MHO\,2424\,A and MHO\,2424\,B form one flow which is
driven by the red source obj.\,E. This object is faint in {\it JHK} but clearly
detected in IRAC. If this is the driving source then the flow would have an
s-shape, indicating the source is a binary. In particular the two knots which
are symmetric and very close (4\arcsec) to source, seem to favour this
explanation.  The total length of the flow would then be about 0.8\arcmin\
(0.6\,pc). However, we cannot rule out that some of the H$_2$ emission, in
particular from MHO\,2424\,A, is driven by the bright EGO (obj.\,A).

\subsubsection*{MHO\,2425}


MHO\,2425 is the a bright bow-shaped emission line object. It appears to form an
open bow pointing towards the NE. As there is no chain of H$_2$ emission (or
other indications such as a counter flow), it is not possible to trace its
source. It could be any of the young sources (obj.\,A\,--\,D). In each case the
flow length between the shock and the source is about 0.5\,pc.

\subsubsection*{MHO\,2426}


The object MHO\,2426\,A consists of two faint knots pointing away from the main
region of outflow driving candidates. In particular the faint EGO (obj.\,B)
seems to be aligned with the flow axis. Furthermore, MHO\,2426\,B could be part
of this flow, as it is aligned with the axis (but see below in the next
subsection). In any case, we would just detect one side of the flow, which would
have a length of 0.5\,pc (0.25\,pc without MHO\,2426\,B). 

\subsubsection*{MHOs\,2426\,B and MHO\,2427}


These three H$_2$ knots are all situated about half a parsec to the
north-north-east of the cluster. If MHO\,2426\,B is not part of a flow with
MHO\,2426\,A, then these three objects could form an independent outflow, as
they are aligned on a common axis with a length of 0.6\,pc. However, there is no
obvious source candidate (red GPS or {\it Spitzer} source) along this axis.
Furthermore, for each H$_2$ object individually, there is no source candidate.
There are other H$_2$ emission features to the West of the cluster. These are
all associated with the HII region and apparent cloud edges in the 8\,$\mu$m
{\it Spitzer} image. Hence all these features are very likely to be
fluorescently excited by either the more massive young stars in the cluster, the
central star of the HII region, or the general UV radiation field. 

\subsubsection*{MHO\,2428}


MHO\,2428 is a single H$_2$ emission knot near MHO\,2423\,B. Based on the
position and the clear flow structure of MHO\,2423, this object seems unrelated.
The most likely source might be the faint EGO (obj.\,B), which is 0.15\,pc away.
However, obj.\,B is also the best candidate source for MHO\,2426\,A (and
MHO\,2426\,B).

\begin{table*}

\caption{\label{sources} IRAC photometry of the possible outflow driving sources
in the Mercer\,14 region. Their coordinates as well as IRAC magnitudes are
listed. Magnitudes for obj.\,C and obj.\,E are taken directly from the GLIMPSE
point source catalogue. Values for the other objects are determined by us (see
Sect.\,\ref{spphot}). Note that typical uncertainties are about 0.2\,mag. We do
not list the K-band excess sources identified in the NIR data which could in
principle drive outflows as well. }   

\begin{center}
\begin{tabular}{lrrrrrrl}
obj. & RA & DEC & I1 & I2 & I3 & I4 & Notes\\
& (J2000) & (J2000) & [mag] & [mag] & [mag] & [mag] & \\
A & 18:58:06.5 & +01:36:52 &  9.5 &  8.4 & 8.0 & 10.3 & bright EGO\,G035.13-0.74 \\
B & 18:58:06.4 & +01:37:02 & 10.6 &  9.7 & 9.0 & 11.3 & faint EGO \\
C & 18:58:05.6 & +01:37:01 & 12.1 &  9.0 & 7.4 & 6.8  & bright IRAC source, no GPS \\
D & 18:58:06.4 & +01:37:07 & 13.2 & 10.1 & 8.7 & 9.2  & faint IRAC source, no GPS \\
E & 18:58:07.5 & +01:36:41 & 11.2 &  9.8 & 8.9 & 8.4  & IRAC and GPS detection \\
\end{tabular}
\end{center}
\end{table*}

\begin{table*}

\caption{\label{mhotab} Molecular hydrogen photometry of the MHOs in the 
Mercer\,14 region. We list the MHO number, the coordinates in (J2000), the
apparent fluxes in the 1-0\,S(1) line, the flow H$_2$ luminosity (dereddened
with A$_K$\,=\,0.8\,mag and assuming 10\,\% of the H$_2$ flux is in the
1-0\,S(1) line), possible flow driving sources (from Table\,\ref{sources}) and
length in pc. A '?' in the last column indicates that the source identification
is unclear.}   

\begin{center}
\begin{tabular}{lllrll}
MHO & RA & DEC & F[1-0\,S(1)] & L[H$_2$] & Possible sources\\
& (J2000) & (J2000) & [W/m$^2$] & [L$_\odot$] & and length \\
2423\,A & 18:58:05.9 & +01:37:20 &  35$\times$10$^{-18}$ & 0.14  & obj.\,D, 0.75\,pc \\
2423\,B & 18:58:07.2 & +01:36:52 &  36$\times$10$^{-18}$ & 0.15  & obj.\,D, 0.75\,pc  \\
2424\,A & 18:58:06.5 & +01:36:52 &  32$\times$10$^{-18}$ & 0.13  & obj.\,A, 0.15\,pc, ? \newline obj.\,E, 0.6\,pc \\
2424\,B & 18:58:08.0 & +01:36:33 &  17$\times$10$^{-18}$ & 0.070 & obj.\,E, 0.6\,pc \\
2425    & 18:58:07.6 & +01:37:30 &  19$\times$10$^{-18}$ & 0.078 & obj.\,C, 0.5\,pc, ? \\
2426\,A & 18:58:06.7 & +01:37:15 & 4.5$\times$10$^{-18}$ & 0.018 & obj.\,B, 0.25\,pc, ? \\
2426\,B & 18:58:07.0 & +01:37:45 & 6.7$\times$10$^{-18}$ & 0.027 & obj.\,B, 0.5\,pc, ? \\
2427\,A & 18:58:08.5 & +01:37:32 & 6.2$\times$10$^{-18}$ & 0.026 & ? \\
2427\,B & 18:58:09.9 & +01:37:22 & 4.7$\times$10$^{-18}$ & 0.019 & ? \\
2428    & 18:58:07.3 & +01:36:58 & 5.0$\times$10$^{-18}$ & 0.020 & obj.\,B, 0.15\,pc, ? \\
\end{tabular}
\end{center}
\end{table*}


\begin{figure*}
\centering
\includegraphics[width=8.5cm]{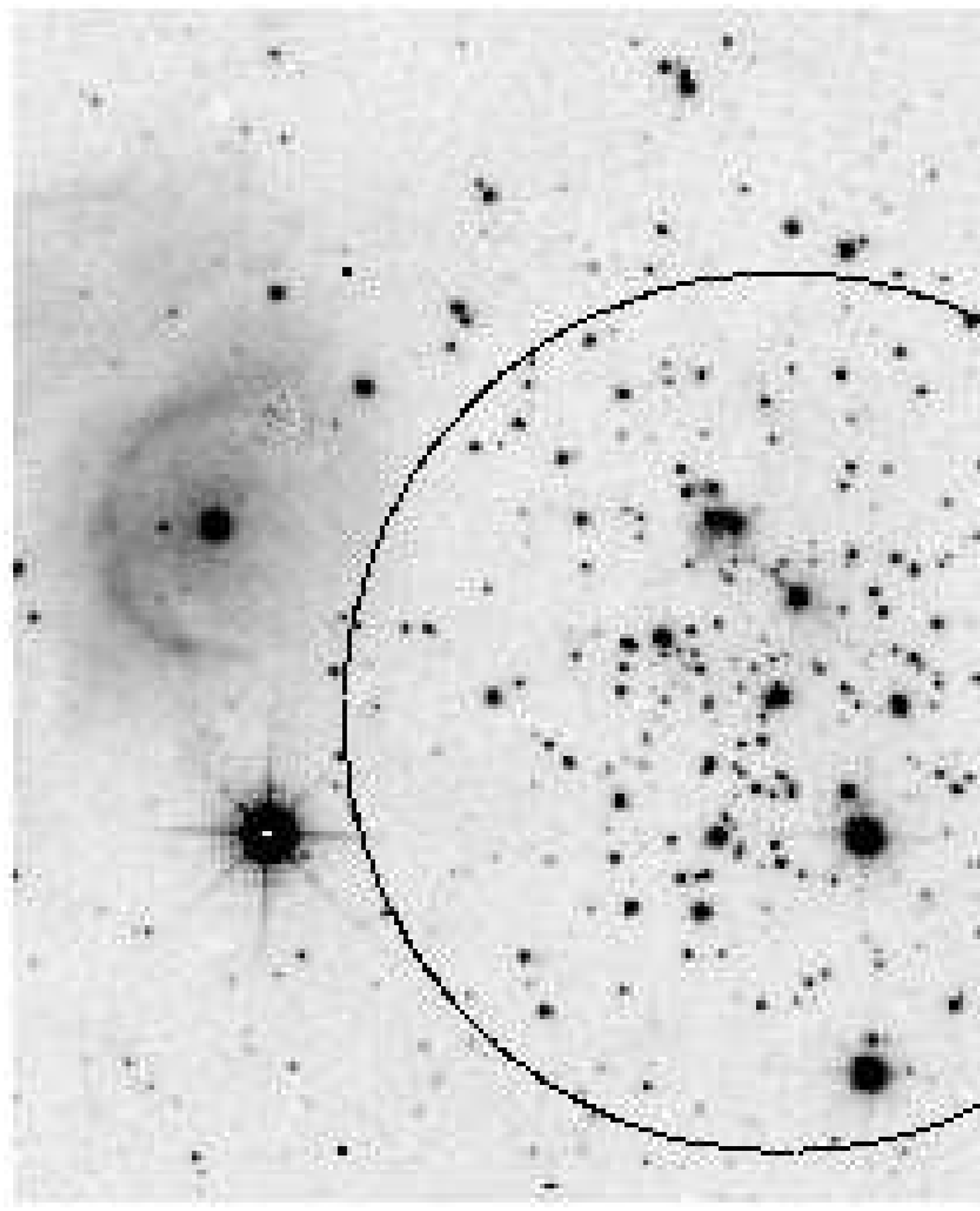} \hfill
\includegraphics[width=8.5cm]{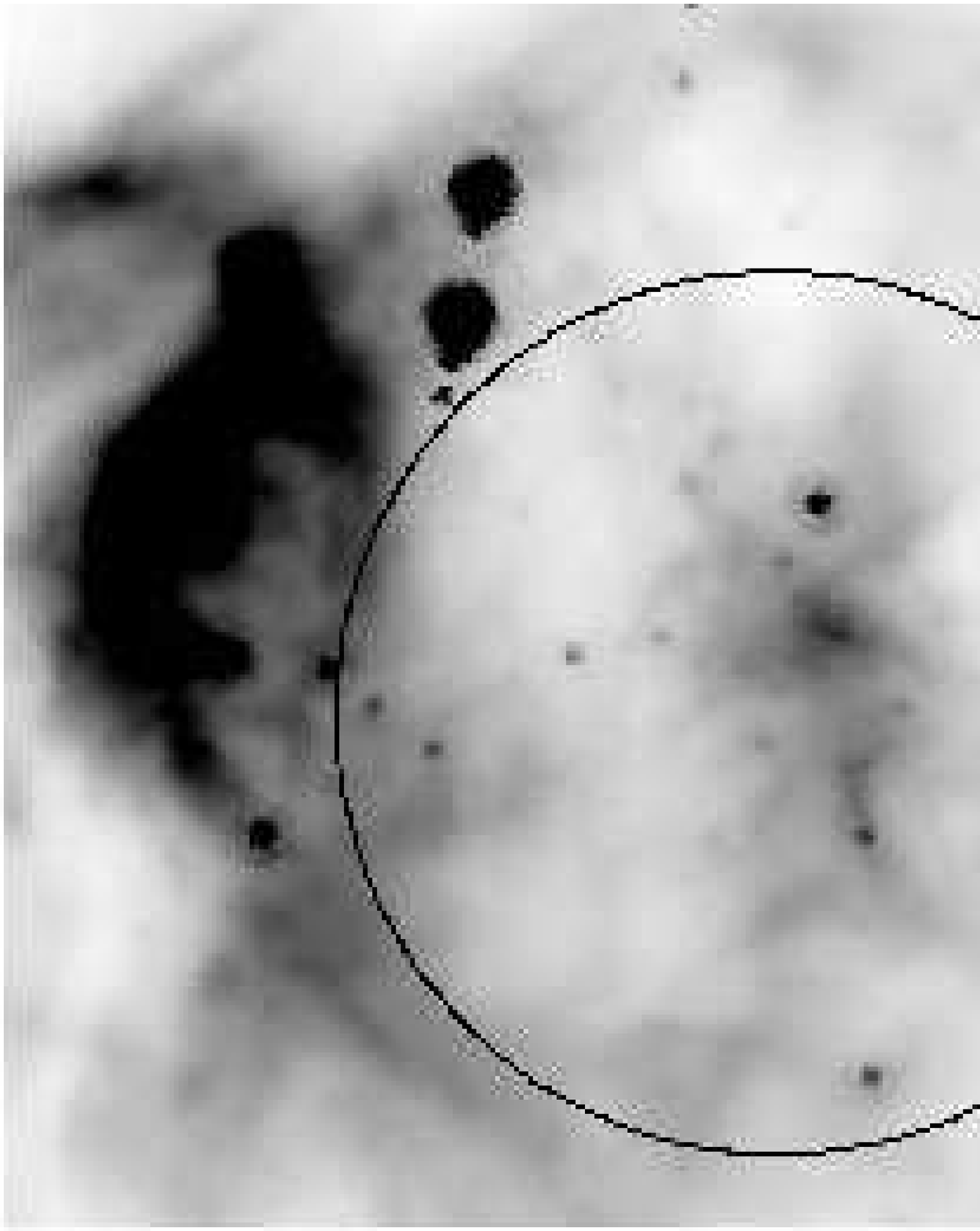}

\caption{\label{k8mu} Gray scale representation of the GPS K-band continuum
(left) and the {\it Spitzer} IRAC 8\,$\mu$m image (right) of the area around
Mercer\,14. The large circle indicates the 'cluster area' centred on
RA\,=\,18:58:06.6 and DEC\,=\,+01:37:03 (J2000) with a radius of 0.9\arcmin.
North is to the top and East to the left in the image.} 

\end{figure*}

\subsection{Cluster properties}
\label{clusterproperties}

Using the decontaminated photometry of the cluster region we plot colour-colour
and colour-magnitude diagrams for the high probability cluster members (see
Fig.\,\ref{mercer14_decon}). We over plot isochrones to determine the cluster
parameters. Given the large distance of 2.5\,kpc, we expect to detect relatively
massive young stars (above about 1\,M$_\odot$) and no lower mass stars in our
NIR data. We thus use isochrones from Girardi et al. \citeyearpar{2002A&A...391..195G},
which provide data for more massive stars. 

The distribution of high probability members in the CMD and in particular the
CCD shows that the cluster suffers from a high amount of extinction, which to a
large extent is intrinsic to the cluster. Hence, the differential reddening is
very high, caused by the youth of Mercer\,14. This is also indicated by the
distribution of the 8\,$\mu$m emission as seen by {\it Spitzer} (right panel
of Fig.\,\ref{k8mu}). There one can clearly see that the cluster coincides with
a 8\,$\mu$m emission peak. There is no correlation of a stars colour with
distance to the emission peak. This indicates that the feature is not foreground
to the cluster. 

We use the CCD and an isochrone for 4\,Myrs (the lowest age available from
Girardi et al. \citeyearpar{2002A&A...391..195G}) to estimate the minimum extinction
shown by a high probability cluster member. This minimum is likely to be the
foreground extinction. Using an extinction law of A$_K$/A$_J$\,=\,0.382 and
A$_K$/A$_H$\,=\,0.612 (based on Mathis \citeyearpar{1990ARA&A..28...37M} and Froebrich
et al. \citeyearpar{2010MNRAS.409.1281F}) we find a minimum foreground extinction of
A$_K$\,=\,0.8\,mag, or A$_V$\,=\,7.4\,mag. This is in good agreement to the
2MASS based extinction maps by Rowles \& Froebrich \citeyearpar{2009MNRAS.395.1640R}.
Their highest resolution map has 0.9\arcmin\ resolution at the cluster position
and determined an optical extinction of 5.9\,mag. The differential reddening,
indicated by the scatter of sources along the reddening band is at least of the
same order of magnitude (A$_K$\,=\,0.8\,mag) as the foreground extinction. 

All high probability members appear in the bottom half of the reddening band in
the CCD, in agreement with the fact that we only detect early spectral type
sources and no low-mass stars in the cluster. A fraction of about one third of
cluster members is found below the reddening band, indicating K-band excess
emission from a disk. This fraction of objects with disk is likely to be a lower
limit and indicates an age of less than 4\,Myrs (Lada \& Lada
\citeyearpar{2003ARA&A..41...57L}). This is also in agreement with the detected outflow
activity and indicates a very low age for the cluster, potentially much less
than the 4\,Myrs used for the isochrone in Fig.\,\ref{mercer14_decon}. 

The three brightest high probability cluster members ($P_{\rm ccm} > 50$\,\%)
have an apparent K-band brightness of about 9.5\,mag. This corresponds to
roughly a 20\,M$_\odot$ mass star. According to Weidner \& Kroupa
\citeyearpar{2006MNRAS.365.1333W} this maximum stellar mass indicates a total mass of
the cluster of about 500\,M$_\odot$. 

\subsection{The nearby HII region}\label{hiiregion}

About 1.2\arcmin\ (0.9\,pc) East of the cluster one can identify a bright star
(18:58:10.5, +01:36:57 (J2000)) within an almost semi-circularly shaped emission
region (see Fig.\,\ref{h2k_image} and left panel of Fig.\,\ref{k8mu}). The
object is listed as planetary nebula PN\,G035.1-00.7 in the SIMBAD database.
Judging by the appearance this seems to be a mis-classification and the object
appears to be an HII region. With an apparent radius of 14\arcsec\ (0.17\,pc)
this corresponds to a compact HII region. The western edge of the region does
not show any H$_2$ emission, possibly because there is no material present to be
excited by the central star, as it has been cleared to low density by the bright
cluster stars in the vicinity.

The central star of the HII region is partly saturated in the UKIDSS GPS data,
we hence use the 2MASS fluxes (J\,=\,10.51\,mag, H\,=\,9.88\,mag,
K\,=\,9.50\,mag) to investigate its nature. We de-redden the star to place it
onto the main sequence in the model isochrones by Girardi et al.
\citeyearpar{2002A&A...391..195G}. With our applied extinction law we require 0.78\,mag
of K-band extinction. This agrees well with the minimum reddening found for the
cluster stars (A$_K$\,=\,0.8\,mag) and thus places the HII region at the same
distance as the cluster. Using the cluster distance, reddening and apparent
magnitudes, we thus find that the central star of the HII region has about 20
solar masses, similar to the brightest cluster members. Hence its spectral type
is about O9 or B0. 

Given the properties of the central star, we can clearly infer that the object
is miss-classified as planetary nebula and indeed an HII region.


\section{Conclusions}\label{conclusions}

We investigate the nature of the young embedded cluster Mercer\,14 utilising
near infrared {\it JHK} photometry from the UKIDSS GPS and narrow-band imaging
data obtained as part of the UWISH2 survey. The cluster shows clear signs of
currently ongoing star formation, indicated by several detected molecular
hydrogen outflows. Mid-infrared {\it Spitzer} data reveals that there are still
substantial amounts of gas and dust within, or in very close proximity to the
cluster. We identify at least five molecular outflows driven by young, partly
deeply embedded sources in the northern half of the cluster.

We estimate the distance of Mercer\,14 by star counts of blue foreground objects
to its parental molecular cloud and a comparison to the Besancon Galaxy model.
We furthermore use the cluster photometric decontamination algorithm to
determine the number of possible foreground stars per unit area towards the
cluster. Both methods result in a distance estimate of 2.5\,$\pm$\,0.3\,kpc.
This puts the cluster at about the same distance as the massive star forming
G\,35.2\,N (distance of 2.3\,kpc), and hence suggests a physical relation of the
two regions. The distance estimation based of the decontaminated cluster
photometry hence has an uncertainty of about 10\,\%. Potentially this method can
be used to determine the distance for a range of young, embedded clusters.

We perform isochrone fitting in the decontaminated colour-colour and
colour-magnitude diagrams to estimate the cluster parameter. The minimum
extinction to the cluster stars is A$_K$\,=\,0.8\,mag, with a differential
extinction of the same order of magnitude. The colours and magnitudes of the
brightest cluster members correspond to 20 solar mass stars. The nearby compact
HII region also contains a star of this mass. This suggests a total cluster mass
of around 500\,M$_\odot$. The content of young massive stars, the detection of
currently ongoing star formation indicated by jets and outflows, a large
fraction of sources with disks, as well as the high differential reddening
caused by remaining gas and dust within the cluster suggests that Mercer\,14 is
a young (less that 4\,Myrs) intermediate mass cluster.

 %
 %

\section*{acknowledgments}

The United Kingdom Infrared Telescope is operated by the Joint Astronomy Centre
on behalf of the Science and Technology Facilities Council of the U.K. The data
reported here were obtained as part of the UKIRT Service Program. This research
has made use of the WEBDA database, operated at the Institute for Astronomy of
the University of Vienna.

\clearpage
\newpage
\begin{appendix}

\section{Detailed images of H$_2$ flows}\label{h2details}

\begin{figure*}
\centering
\includegraphics[width=10cm]{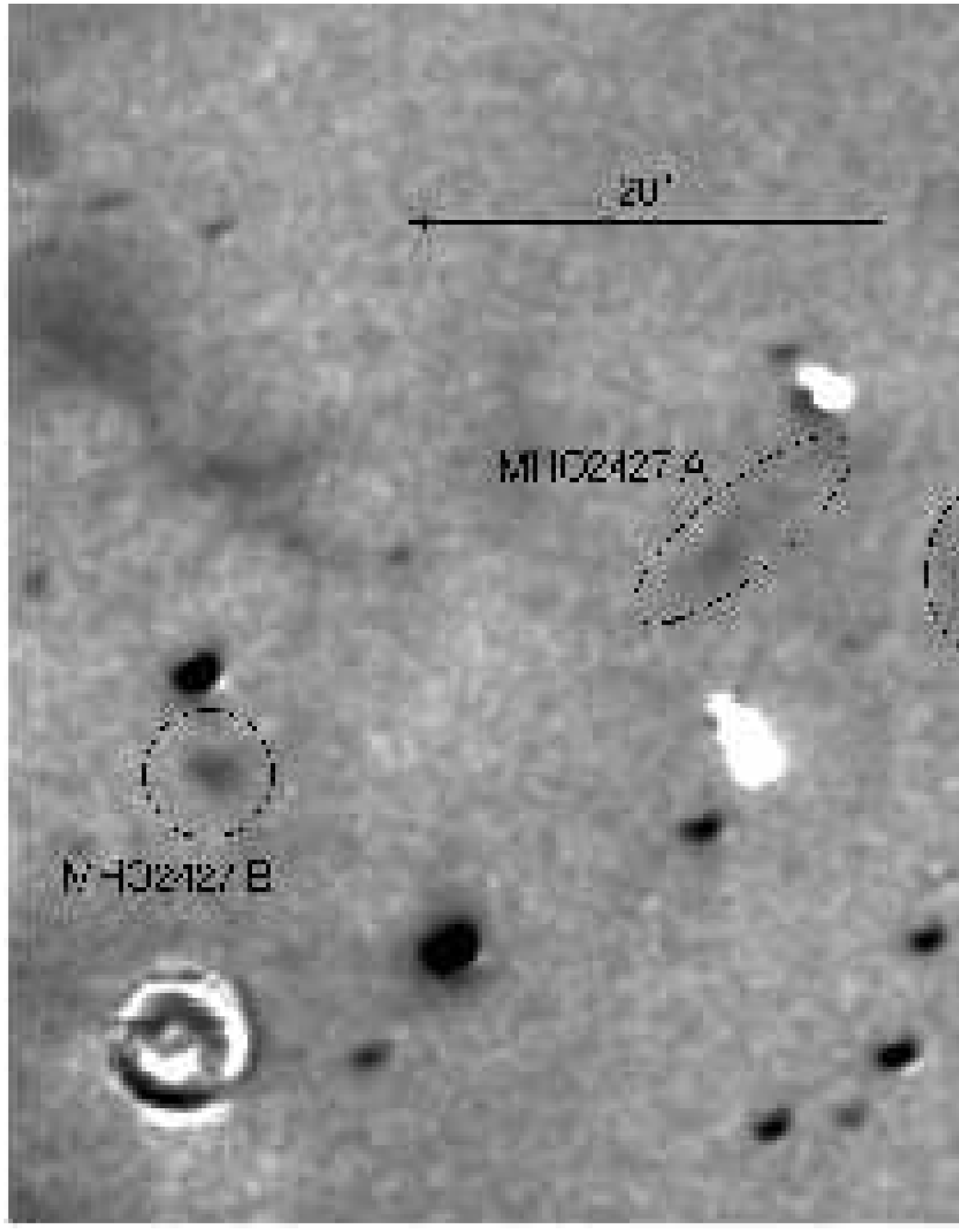}
\caption{\label{app1} Gray scale representation of the continuum subtracted
H$_2$ image of the outflows near Mercer\,14. Here we show the northern part of
the region. North is up, East to the left. } 
\end{figure*}

 %
 %
 %

\clearpage
\newpage

\begin{figure*}
\centering
\includegraphics[width=10cm]{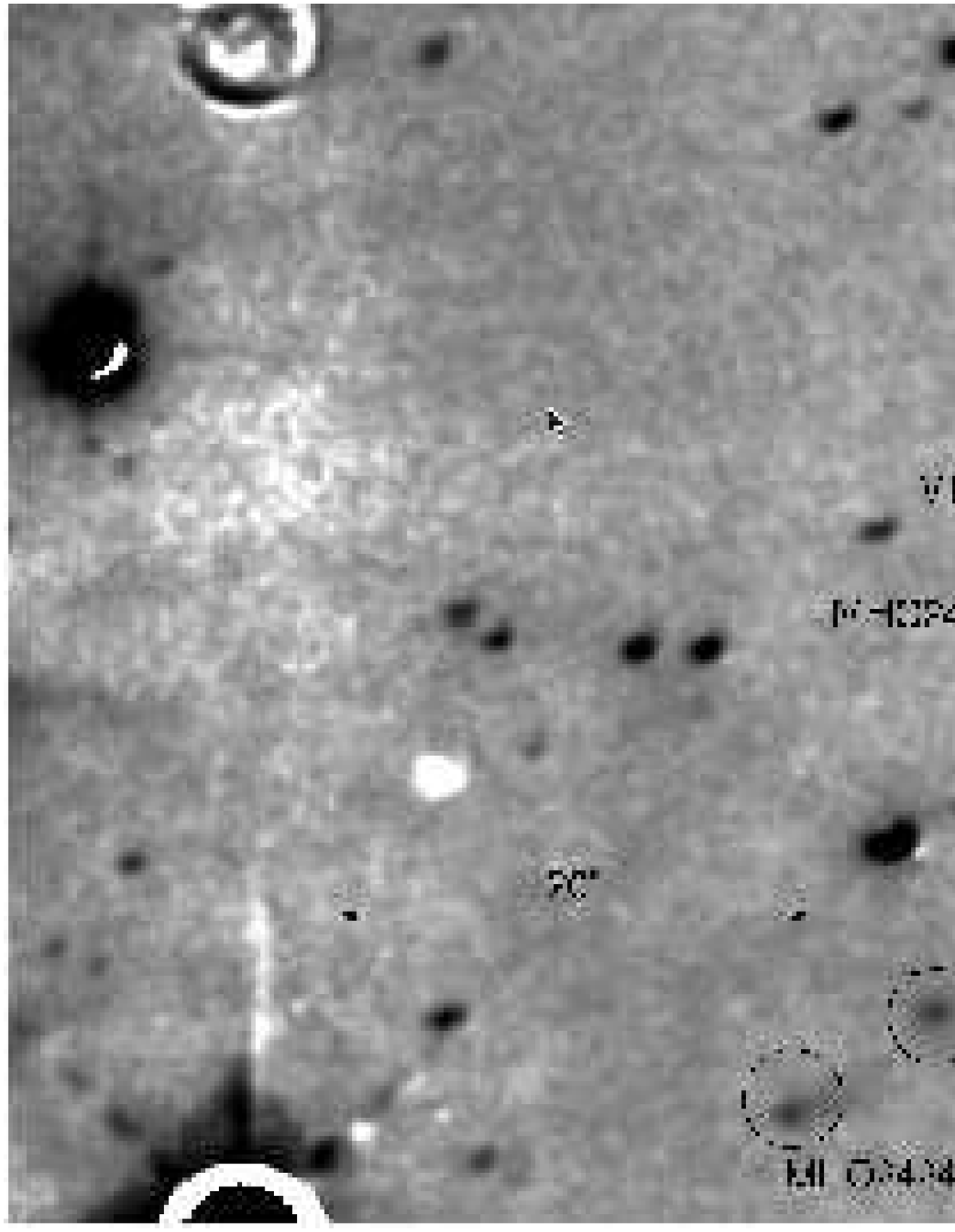}
\caption{\label{app3} Gray scale representation of the continuum subtracted
H$_2$ image of the outflows near Mercer\,14. Here we show the southern part of
the region. North is up, East to the left. } 
\end{figure*}

\end{appendix}

\label{lastpage}

\end{document}